\documentclass{article}
\usepackage{spconf,amsmath,graphicx}

\usepackage{url}
\usepackage{graphicx}
\usepackage{mathtools}
\usepackage{makecell}
\usepackage{color, soul}
\usepackage{multirow}
\usepackage{subfigure}
\usepackage{stfloats}
\usepackage{enumitem}
\usepackage{array}

\usepackage{xcolor,colortbl}
\definecolor{Gray}{gray}{0.90}
\definecolor{Gray1}{gray}{0.95}

\usepackage{acro}

\DeclareAcronym{asic}{short = ASIC , long  = Application-Specific Integrated Circuit}
\DeclareAcronym{amt}{short = AMT , long  = Adaptive Multiple Transform}

\DeclareAcronym{brams}{short = BRAMS , long  = Block RAMS}

\DeclareAcronym{dct}{short = DCT , long  = Discrete Cosine Transform}
\DeclareAcronym{dst}{short = DST , long  = Discrete Sine Transform }
\DeclareAcronym{dc}{short = DC , long  = Design Compiler}

\DeclareAcronym{fpga}{short = FPGA , long  = Field-Programmable Gate Array}
\DeclareAcronym{fpgas}{short = FPGAs , long  = Field-Programmable Gate Array}
\DeclareAcronym{hevc}{short = HEVC , long  = High Efficient Video Coding}

\DeclareAcronym{ietr}{short = IETR , long  = Institut d'\'Electronique et de T\'el\'ecommunications de Rennes}
\DeclareAcronym{idst}{short = IDST , long  = Inverse DST }
\DeclareAcronym{idct-ii}{short = IDCT-II , long  = Inverse DCT-II}

\DeclareAcronym{jvet}{short = JVET , long  = Joint Video Experts Team}

\DeclareAcronym{mts}{short = MTS , long  = Multiple Transform Selection }
\DeclareAcronym{mpeg}{short = MPEG , long  = Motion Picture Experts Group}

\DeclareAcronym{ram}{short = RAM , long  = Random-Access Memory}
\DeclareAcronym{rom}{short = ROM , long  = Read-Only Memory}
\DeclareAcronym{rm}{short = RM , long  = Regular Multiplier}
\DeclareAcronym{mcm}{short = MCM , long  = Multiple Constant Multiplier}

\DeclareAcronym{tsmc}{short = TSMC , long  = Taiwan Semiconductor Manufacturing Company}

\DeclareAcronym{vvc}{short = VVC , long  = Versatile Video Coding}
\DeclareAcronym{vceg}{short = VCEG , long  = Video Coding Experts Group}

\title{Asymmetric Gained Deep Image Compression With Continuous Rate Adaptation}
\name{Ze Cui$^{\dagger}$, Jing Wang$^{\star \dagger}$, Shangyin Gao$^{\dagger}$, Tiansheng Guo$^{\dagger}$, Yihui Feng$^\dagger$ and Bo Bai$^{\dagger}$,}
\address{$^\dagger$ Huawei, China. \\ 
	 \\$\star$E-mail: wangjing215@huawei.com
}

\begin{document}
\ninept
\maketitle
\begin{abstract}
With the development of deep learning techniques, the combination of deep learning with image compression has drawn lots of attention. Recently, learned image compression methods had exceeded their classical counterparts in terms of rate-distortion performance. However, continuous rate adaptation remains an open question. Some learned image compression methods use multiple networks for multiple rates, while others use one single model at the expense of computational complexity increase and performance degradation. In this paper, we propose a continuously rate adjustable learned image compression framework, Asymmetric Gained Variational Autoencoder (AG-VAE). AG-VAE utilizes a pair of gain units to achieve discrete rate adaptation in one single model with a negligible additional computation. Then, by using exponential interpolation, continuous rate adaptation is achieved without compromising performance. Besides, we propose the asymmetric Gaussian entropy model for more accurate entropy estimation. Exhaustive experiments show that our method achieves comparable quantitative performance with SOTA learned image compression methods and better qualitative performance than classical image codecs. In the ablation study, we confirm the usefulness and superiority of gain units and the asymmetric Gaussian entropy model.
\end{abstract}

\acresetall

\section{Introduction}

Image compression is one of the most fundamental and valuable problems in image processing and computer vision. In the last decades, many researchers have worked for the development and optimization of the classical image compression codecs, such as JPEG \cite{Grepory1992}, JPEG2000 \cite{Majid2012} and BPG \cite{Fabrice2014}. To remove redundancy within images, basic modules of the classical codes, including transform coding, entropy coding and quantization, have been sophistically designed and applied. Since these modules are artificially designed and optimized separately, it is not easy to obtain an optimal solution for different evaluation indicators.
\begin{figure}
	\centering
	\vspace{-0.1cm}
	\includegraphics[width=8.4cm]{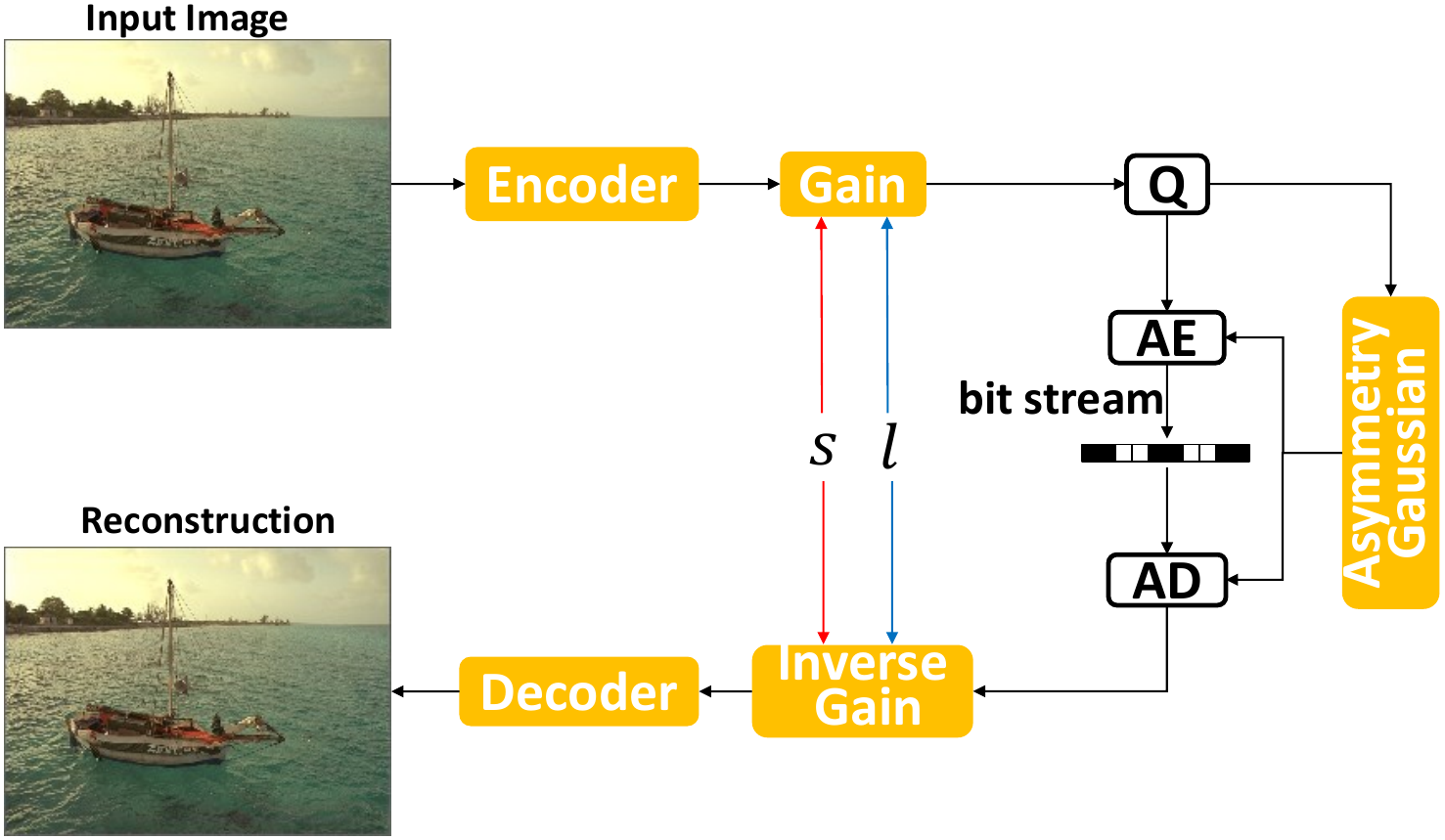}
	\caption{AG-VAE framework. We achieve rate adaptation by inserting a gain unit after encoder and an inverse gain unit before decoder. The bit rate could be adjusted continuously with the change of the gain vector index $s$ and the interpolation coefficient $l$. The asymmetric Gaussian entropy model estimates entropy of the gained and quantized latent representation accurately.}
	\setlength{\abovecaptionskip}{-0.2cm}
	\vspace{-0.70cm}
	\label{fig:framework}
\end{figure}

Some VAE-based image compression methods need to train multiple fixed-rate models to realize rate adaption, each model for one rate. Therefore, the training cost and memory requirement increase dramatically with the growth and refinement of the desired rate range. Instead of using multiply models, some other methods achieve the rate adaptation using one single model. The RNN-based schemes \cite{Toderici2016, Toderici2017, Johnston2018} encode the input image progressively,  but they suffer from bad R-D performance. The conditional autoencoder \cite{Choi2019, Yang2019} incorporates fully connected layers into the convolution unit to achieve discrete rate adaptation while increasing the network's computational complexity and memory requirement. Mixed bin sizes \cite{Choi2019} are introduced to extend the range coverage from finite discrete points to a broad rate range, but they induce R-D performance degradation. Bottleneck scaling scheme \cite{TheisLossy2017, Akbari2019} ignores compatibility between autoencoder and scaling parameters and has a poor performance in low bit rate range. Although providing feasible solutions to rate adaptation in a single model, the methods mentioned above have various practical problems such as performance degradation, computational complexity increase and memory increase.

As shown in Figure~\ref{fig:framework}, we propose a novel image compression framework, AG-VAE. It can continuously adjust the bit rate in one single model and achieves comparable R-D performance with SOTA learned image compression methods in quantitative metrics and qualitative visual quality. Based on the unevenness of channel redundancy, we design a plug-and-play variable-rate block, gain unit. By simple channel-wise multiplication, the gain unit rescales the latent representation. The degree of information loss is then controlled in the quantization process. 

We address two critical challenges of the proposed framework. First, the inverse-gain unit is introduced to avoid performance degradation. Second, we study the reconstruction assumption of gain units to deduce the exponent interpolation formula, enabling continuous rate adaptation without extra training. To avoid entropy estimation error for the samples with the asymmetric distribution, we also introduce the asymmetric Gaussian entropy model to achieve good R-D performance. To demonstrate the universality of gain units, we further integrate them into other backbone architectures \cite{BallVariational2018,MinnenJoint2018}. Besides, we also compare gain units with previous rate adaptation methods from additional computation and performance degradation. 

Our contributions can be summarized as follows:
\begin{itemize}[itemsep=0pt,topsep=0.1pt,listparindent=0.1pt]
	\item We introduce gain units to achieve discrete rate adaptation in one single model. With negligible additional computational cost, our method has a similar performance with SOTA learned image compression method.
	\item We propose the exponent interpolation, which can generate gain vectors at the arbitrary bit rate. The exponent interpolation formula extends the rate's coverage from finite discrete points to a broad continuous range without an extra training process.
	\item Gain units with exponent interpolation can be easily generalized to all VAE-based image compression methods while avoiding performance degradation.
	\item We propose the asymmetric Gaussian entropy model to achieve more accurate entropy estimation. Less bit rate is required to reach the same distortion level.
\end{itemize}

\section{Related Works}
\noindent \textbf{Learned Image Compression.} The VAE-based framework could be counted as a nonlinear transforming coding model \cite{balle2016,balle2017}. The transforming process could be mainly divided into four parts: The encoder that maps an image $x$ into a latent representation, $y=f_{\theta}\left ( x \right )$; The quantizer that transforms the latent representation into the discrete values, $\hat{y}=Q\left ( y \right )$; The entropy model that estimates the distribution of $\hat{y}$ to get the minimum rate achievable with lossless entropy source coding \cite{Thomas2012}, $R_{\varphi}\left( \hat{y} \right)$; And the decoder that transforms the quantized latent representation to the image, $\hat{x}=g_{\phi }\left ( \hat{y} \right )$. The entire framework can be trained jointly by optimizing the following loss function as:
\begin{equation}
	\underset{\theta,\phi,\varphi}{min} \quad R_{\varphi}\left ( Q\left ( f_{\theta}\left ( x \right ) \right ) \right ) + \beta \cdot D\left(x, g_{\phi }\left ( Q\left ( f_{\theta}\left ( x \right ) \right ) \right ) \right),
	\setlength\abovedisplayskip{3pt}
	\setlength\belowdisplayskip{3pt}
	\label{eq:basic}
\end{equation}
where $ R_{\varphi}\left ( \cdot  \right )$ represents the expected code length (bit rate) of the quantized latent representation and $ D\left ( \cdot  \right )$ measures the distortion between the input image and the reconstructed image. The Lagrange multiplier $\beta$ is a constant in the training process to specify the R-D tradeoff of the trained model \cite{Antonio1998}. Therefore, the VAE-based image compression methods need to use multiple fixed-rate models trained under different $\beta$ to adjust the different compression performance of images. However, the multi-model scheme only realizes variable rates in several discrete points of the R-D curve, while memory consumption increases proportionally.

\noindent \textbf{Rate Adaptation Methods.} The first variable-rate learned image compression was proposed by Toderici et al. \cite{Toderici2016}. Instead of autoencoder structure, they adopt convolutional LSTM networks. The network is only trained once and can progressively transmit bits. The more bits are sent, the more accurate the image reconstruction is. Subsequently, the LSTM-based scheme was widely adopted, and new techniques were absorbed in it, such as residual scale reconstruction, better entropy coding, and spatial adaptive bit rates \cite{Toderici2017,Johnston2018}. However, the LSTM-based schemes can not outperform JPEG2000 \cite{Majid2012} in terms of R-D performance and can not achieve continuous variable-rate adaptation. The LSTM network needs to be inferred multiple times to reconstruct a high-quality image, which is time-consuming and impractical for real-world applications.

Choi et al. \cite{Choi2019} proposed a variable rate image compression framework with a conditional autoencoder, which incorporated fully connected networks into the convolution unit and adjusted compression performance with the Lagrange multiplier. Mixed bin sizes were introduced to control quantization loss and finetune bit rate in \cite{Choi2019}. However, additional fully connected layers of the conditional convolution increase the computational complexity and memory of the network. Besides, the adjustment of the bin size influences the R-D performance to some extent. It also causes the dilemma of selecting the best combination of the Lagrange multiplier and quantization size in the intersection of adjacent coarse-adjusting coverages. Yang et al. \cite{Yang2019} proposed a modulated autoencoder to realize rate adaptation in several discrete points of the R-D curve. Similar to conditional autoencoder \cite{Choi2019}, the modulated network introduced fully-connected layers into the autoencoder, which also caused the increase of memory and computation. 

Thesis et al. \cite{TheisLossy2017} first trained the autoencoder networks at a high bit rate. The pre-trained autoencoder was then fixed and incorporated with scale parameters to achieve rate adaptation. Nevertheless, the incompatibility between autoencoder and scaling parameters led to performance degradation, especially in the R-D curve's low-rate segmentations. Akabari et al. \cite{Akbari2019} proposed a stochastic rounding-based quantization scheme and replaced the loss term with rate estimation of the loss function to enable a single model to operate at different bit rates. However, the alteration of loss function made its performance in PSNR much lower than BPG \cite{Fabrice2014}. Compared to the closely related bottleneck scaling methods \cite{Fabrice2014,TheisLossy2017}, the proposed gain units provide more insights on strengthening autoencoder's suitability, gain unit, and inverse gain unit to avoid R-D performance degradation. Based on the reconstruction assumption, we deduce the exponent interpolation formula to achieve continuous rate adaptation in the whole R-D curve.

By incorporating the proposed gain units and a series of optimization schemes \cite{zhang2019,Ziv1985,Zamir1992,Oord2016}, we have participated in the Workshop and Challenge on Learned Image Compression 2020 (CLIC2020) \cite{CLIC2020} and achieved good performance for low bit-rate image compression task \cite{Guo2020}. In this paper, we will introduce the motivations and principles of the gain units with exponential interpolation in detail, which can be easily generalized to all VAE-based image compression methods to achieve continuous rate adaptation.

\noindent \textbf{Entropy Estimation Model.} To obtain accurate entropy estimation of the latent representation, Ball$\acute{e}$ et al. \cite{BallVariational2018} firstly proposed a zero-mean to model the latent representation. Subsequently, Minnen et al. \cite{BallVariational2018} proposed to estimates the distribution of the latent representation and hyperprior with a mean-and-scale Gaussian entropy model and a non-parametric, fully factorized density model respectively to get the minimum rate achievable with lossless entropy source coding \cite{Thomas2012}, which was still in use by the current learned image compressions methods \cite{MinnenJoint2018,Choi2019,Mentzer2018,ZhouLei2019}. However, the symmetric Gaussian entropy model has insufficient degrees of freedom and may induce large estimation errors for natural images with other distributions.

\section{Proposed Method}
In this section, we present our proposed image compression framework AG-VAE, as shown in Figure~\ref{fig:framework}. First, we introduce the principles of the gain units for discrete rate adaption. Then, we depict how the exponent interpolation formula enables the gain units to achieve continuous rate adaption. Furthermore, we extend the gain unit to hyperprior to save more bit rates. Finally, we discuss the effectiveness of the asymmetric Gaussian entropy model.
\subsection{Gain Unit}
\label{section:sec3.1}
\begin{figure}
	\centering
	\vspace{-0.1cm}
	\subfigure{
		\label{fig:channel_contrast}
		\includegraphics[width=3.8cm]{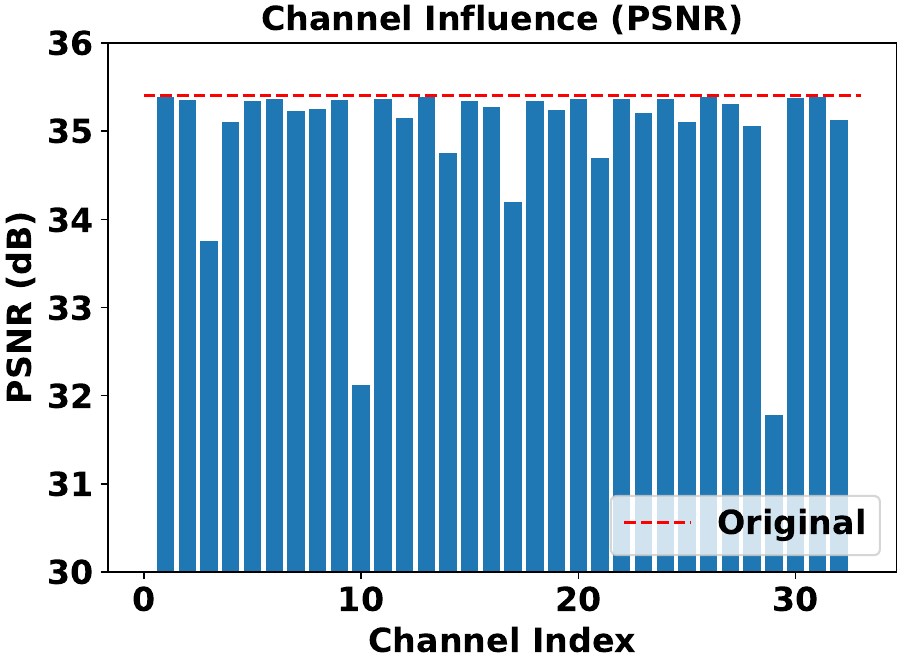}
	}
	\hspace{-3.0mm}
	\subfigure{
		\label{fig:scale_contrast}
		\includegraphics[width=3.8cm]{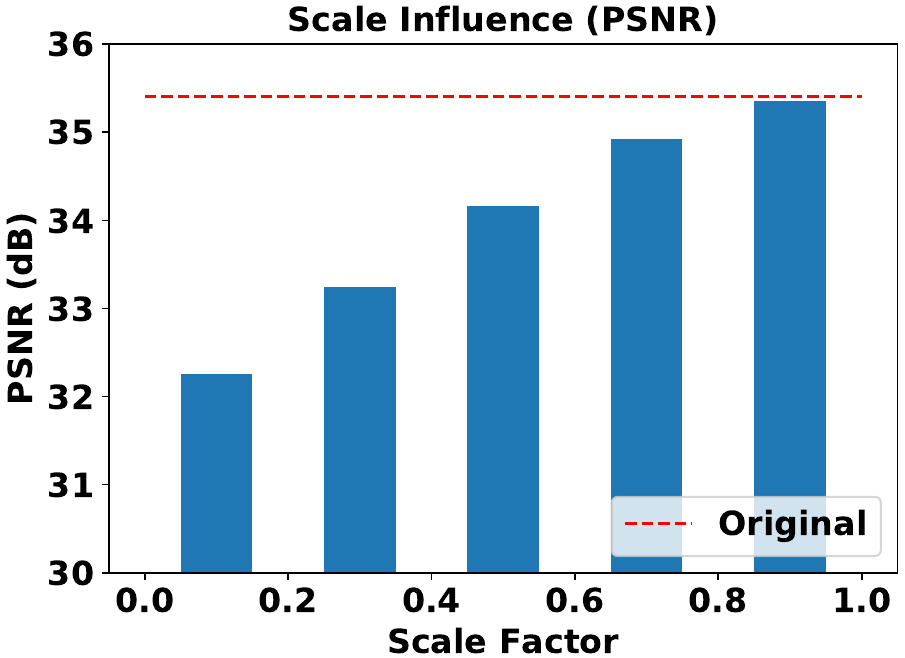}
	}
	\caption{Illustration of channel influences on reconstruction distortion. Left: PSNR degradation of each channel(channel by channel, the first 32 channels of the quantized feature map). Right: PSNR degradation of one channel with various scale factors.}
	\setlength{\abovecaptionskip}{-0.1cm}
	\vspace{-0.7cm} 
	\label{fig:channel}
\end{figure}

Here, we first conduct a simple experiment to show the channel-wise uneven redundancy in latent representation that widely exists in VAE-based learned image compression frameworks. We taken one image as input of the encoder to obtain its latent representation, denoted as $y \in R^{c \times h \times w}$, where $c, h, w$ represents the number of channels, height, and width of the latent representation respectively. Each channel of $y$ can be denoted as $y_{i} \in R^{h \times w}$, where $i =0,1,\cdots,c-1$ (If not explicitly mentioned, $c$ is 192 in our framework). We take $kodim20$ from Kodak dataset as an example and set the first 32 channels of the latent representation to zero individually. The modified latent representation is then converted back to the RGB domain, and the PSNR degradation of the reconstruction by the absence of different channels is shown in the left part of Figure~\ref{fig:channel}. The Channel-29 is selected as an example due to the worst degradation in the absence experiment, and the corresponding PSNR of the reconstruction under different scale factors is depicted in the right part of Figure~\ref{fig:channel}. With the decrease of the scaling factor, the quality of the reconstruction is also reduced. We can conclude that the channels' importance varies and can be scaled to control the reconstruction quality. However, lots of the learned image compression methods ignore the uneven redundancy between channels and treat them equally in the quantization process\cite{Zhong2020}.

To fully utilize the above-mentioned property and scale the latent representation flexibly, we design the gain unit. The gain unit is made up of a gain matrix $M \in R^{c \times n}$, where $n$ represents the number of gain vectors. The gain vector can be denoted as $m_{s} = \left \{ m_{s,0}, m_{s,1}, \cdots, m_{s,c-1} \right \}$, where $s$ represents the index of the gain vectors in the gain matrix. And $m_{s,i} \in R$ represents the $i$th gain value in the gain vector $m_{s}$ and $i$ ranges from 0 to $c-1$. Each channel is associated with its own scale value. The rescale operation of the latent representation is depicted as:
\begin{equation}
	\bar{y}_{s,i }=y_{i} \times m_{s,i}
	\setlength\abovedisplayskip{3pt}
	\setlength\belowdisplayskip{3pt}
	\label{eq:gain_channel}
\end{equation}

In this way, the quantization loss of the latent representation can be finely adjusted by the gain vector channel-wisely. Therefore, the network is guided to allocate more bit rates for the channels, which influence the reconstruction quality significantly. The calculation process of the gain unit can be described as:
\begin{equation}
	\bar{y}_{s} = G_{\psi} \left(y, s \right) = y \odot m_{s},
	\setlength\abovedisplayskip{3pt}
	\setlength\belowdisplayskip{3pt}
	\label{eq:gain}
\end{equation}
where $\bar{y}_{s}$ is the gained latent representation, $G_{\psi} \left(\cdot \right)$ represents the gain process, and $\odot$ represents channel-wise multiplication in Eq~\ref{eq:gain_channel}. What needs to be mentioned is that the gain matrix is trained jointly with the autoencoder network to ensure compatibility between them.
\subsection{Discrete Variable Rate with Gain Units}
In the VAE-based image compression methods, the quantizer is applied element-wisely to round the latent representation $y$ to the nearest integer. In the Section~\ref{section:sec3.1}, we show that the channel redundancy of the $y$ is uneven. By scaling the $y$ to different intervals channel wisely, the gain unit can adjust the channel redundancy, thus control information loss of the quantization process effectively. The quantization process can be formulated as:
\begin{equation}
	\hat{y}_{s} = Q(\bar{y}_{s}) =  round(\bar{y}_{s}),
	\setlength\abovedisplayskip{3pt}
	\setlength\belowdisplayskip{3pt}
	\label{eq:quantization}
\end{equation}
where $\hat{y}_{s}$ represents the quantized gained latent representation, $Q\left(\cdot \right)$ is the quantization process and $round(\cdot)$ denotes element-wisely rounding operation. 
\begin{figure}
	\centering
	\vspace{-0.1cm}
	\includegraphics[width=8.4cm]{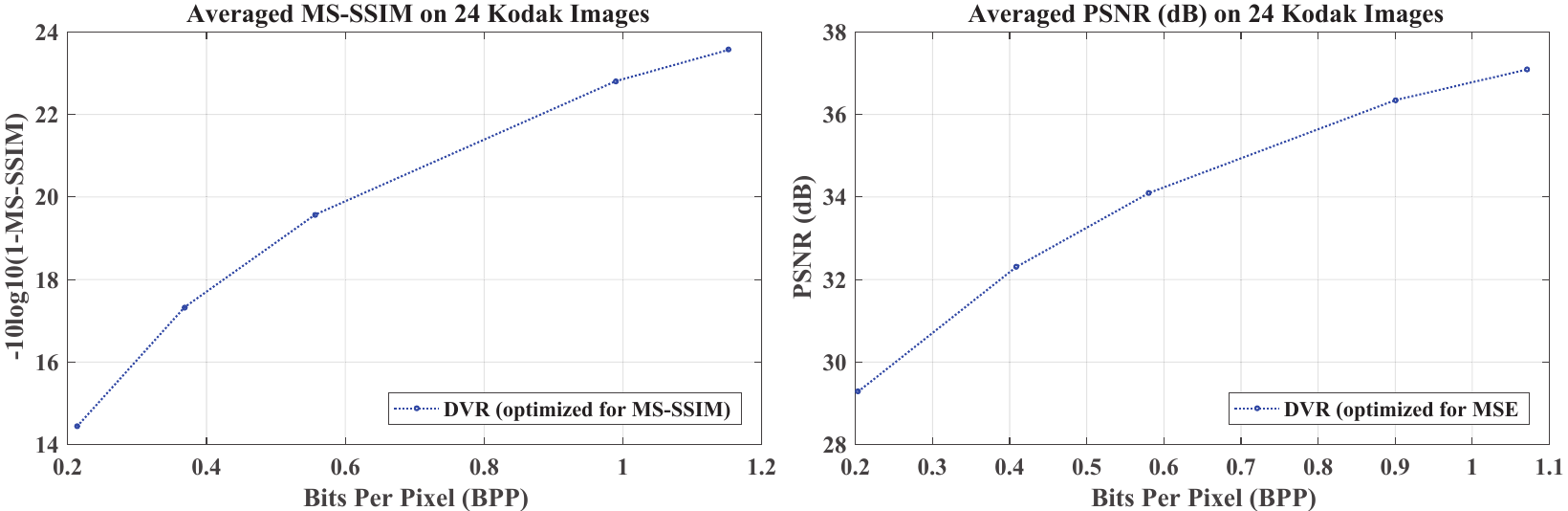}
	\caption{The R-D performance of our DVR method with gain units on 24 Kodak images. In this experiment, we set $n$ to 5 so that it can produce 5 points of the R-D curve in a single model.}
	\setlength{\abovecaptionskip}{-0.2cm}
	\vspace{-0.70cm}
	\label{fig:dvr}
\end{figure} 

Before giving the rescaled and quantized latent representation $\hat{y}_{s}$ to the decoder, an inverse rescale operation needs to been done. This operation is used to map $\hat{y}_{s}$ back to the same numerical intervals as $y$, thus ensuring the reconstruction's correctness. \cite{TheisLossy2017} limits the scale and inverse-scale operation to be strictly reciprocal. However, they ignore that the latent representation can not be mapped to the same numeral intervals due to quantization operation by reciprocal inverse scale operation. Here, we adopt another trainable gain unit before the decoder to adaptively rescale $\hat{y}_{s}$, named as the inverse gain unit. Consequently, the gain matrix and gain vector in the inverse-gain unit are denoted as inverse-gain matrix ${M}'\in R^{c \times n}$ and inverse gain vector ${m}'_{s}= \left \{ {m}'_{s,0}, {m}'_{s,1}, \cdots, {m}'_{s,c-1} \right \},  {m}'_{s,i} \in R$. The inverse-gain process can be represented as:
\begin{equation}
	{y}'_{s} = IG_{\tau}\left( \hat{y}_{s}, s\right)= \hat{y}_{s} \odot {m}'_{s},
	\setlength\abovedisplayskip{3pt}
	\setlength\belowdisplayskip{3pt}
	\label{eq:inverse_gain}
\end{equation}
where $IG_{\tau}\left( \cdot \right)$ represents the inverse gain process. And $\odot$ represents channel-wise multiplication similar with Eq~\ref{eq:gain_channel}.

The inverse gain vector ${m}'_{s}$ and the corresponding gain vector $m_{s}$ always appear in pairs, which could be expressed as $\left \{m_{s}, {m}'_{s} \right \}$. In the training process, each pair of gain vectors $\left \{m_{s}, {m}'_{s} \right \}$ corresponds to a specific Lagrange multiplier $\beta_{s}$ from the predefined finite set of the Lagrange multipliers, $B \in R^{n}$. The gain vector, inverse-gain vector, and Lagrange multiplier are bound together with the subscript $s$. Thus, the loss function of the discrete variable rate (DVR) framework is defined as below:
\begin{equation}
	\begin{split}
		\underset{\theta,\phi,\varphi,\psi}{min} \! \quad \! &\sum_{s=0}^{n-1} \! R_{\varphi}\left (\! Q\left (\! G_{\psi}\left ( \!f_{\theta}\left ( x \right ), \!s \right ) \right ) \right ) \\
		& + \!\beta_{s} \!\cdot \! D\left(x\!, \!g_{\phi }\left (\! IG_{\tau}\left (\! Q\left (\! G\left (\! f_{\theta}\left ( x \right ), \!s \right ) \right ),\! s\right ) \right ) \right),
	\end{split}
	\setlength\abovedisplayskip{3pt}
	\setlength\belowdisplayskip{3pt}
	\label{eq:loss_dvr}
\end{equation}
where $G_{\psi}\left( \cdot \right)$ and $IG_{\tau}\left( \cdot \right)$ represents the gain process and inverse gain process respectively, $ R_{\varphi}\left ( \cdot  \right )$ represents the expected bit rate of the quantized gained latent representation.

In the inference process, we change $s$ to obtain the corresponding gain and inverse-gain vector pair, which could be used to scale the distribution of $y$ and ${y}'_{s}$ respectively. By this means, we can obtain the desired compression performance limited to several discrete points of the R-D curve. The the R-D curve range depends on the number and value of Lagrange multiplier $\beta_{s} \in B$. We denote the VAE-based image compression method with gain units as the discrete variable rate (DVR) method. It can be seen from Figure~\ref{fig:dvr} that the DVR method could achieve rate adaptation among several discrete points of the R-D curve in a single model.

\begin{figure}
	\centering
	\vspace{-0.1cm}
	\includegraphics[width=8.4cm]{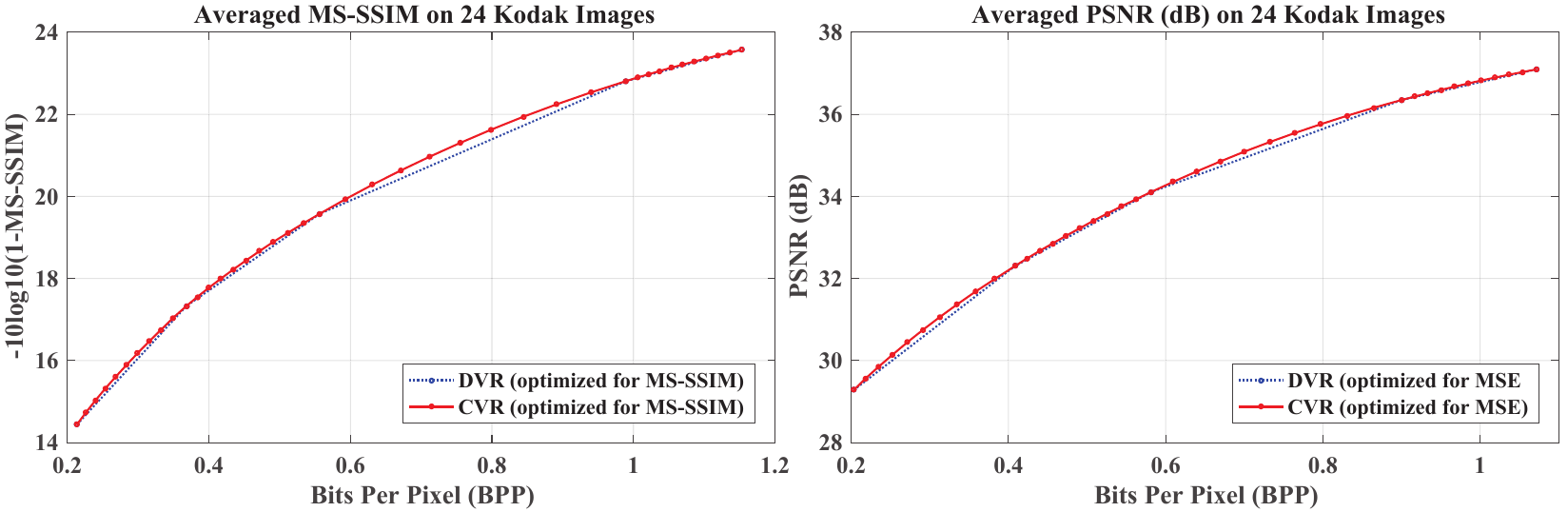}
	\caption{PSNR and MS-SSIM comparison between our DVR method and our CVR method on 24 Kodak images. CVR method owns the same architecture and parameters as DVR method.}
	\setlength{\abovecaptionskip}{-0.2cm}
	\vspace{-0.70cm}
	\label{fig:cvr}
\end{figure} 

\subsection{Exponential Interpolation}
Since we use different gain unit pairs to achieve discrete rate adaptation, continuous one can be achieved by interpolation between gain units. In this section, we derive the exponential interpolation based on the property of gain units. The gain unit pair ensures the numerical intervals of $\hat{y}$ and $y$ to be the same, which can be formulated as:
\begin{equation}
	m_{t}\cdot m_{t}^{'} = m_{r}\cdot m_{r}^{'} = C,
	\setlength\abovedisplayskip{3pt}
	\setlength\belowdisplayskip{3pt}
	\label{eq:reconstruction}
\end{equation}
where $\left \{m_{t}, {m}'_{t} \right \}$ and $\left \{m_{r}, {m}'_{r} \right \}$ ($r, t \in \left [ 0, 1, \cdots, n-1\right ]$) represent the gain vector pairs corresponding to different bit rates, and $C\in R^{c}$ is a constant vector. According to Eq~\ref{eq:reconstruction}, we can derive the exponent interpolation formula as:
\begin{equation}
	\begin{split}
		\left ( m_{r}\!\cdot\! m_{r}^{'} \right )^{l} \!\cdot \!\left  ( m_{t}\cdot m_{t}^{'} \right )^{1-l}\!=\!C, \\
		\left [ \left ( m_{r} \right )^{l}\! \cdot\! \left (m _{t} \right )^{1-l}\right ]\!\cdot \!\left [ \left ( m_{r}^{'} \right )^{l} \!\cdot \! \left (m _{t}^{'} \right )^{1-l}\right ]\!=\!C, \\
		m_{v}\! = \!\left [ \left ( m_{r} \right )^{l}\! \cdot \! \left (m _{t} \right )^{1-l}\right], \! m_{v}^{'}\!=\!\left [ \left ( m_{r}^{'} \right )^{l}\! \cdot\! \left (m _{t}^{'} \right )^{1-l}\right ],
		\setlength\abovedisplayskip{1pt}
		\setlength\belowdisplayskip{1pt}
	\end{split}
	\label{eq:interpolation}
\end{equation}
where  $\left \{m_{v}, {m}'_{v} \right \}$ is the generated gain vector pair and $l \in R$ is an interpolation coefficient, which controls the corresponding bit rate of the generated gain vector pair. Since $l$ is a real number, utilizing the exponent interpolation of the gain vector pairs could achieve an arbitrary bit rate between $t$ and $r$. And when $l$ is equal to 0 or 1, it represents $\left \{m_{t}, {m}'_{t} \right \}$ or $\left \{m_{r}, {m}'_{r} \right \}$ respectively. Without an extra training process and supplementary blocks, we apply the exponent interpolation formula between the adjacent gain vector pairs in the inference process to obtain the Continuously Variable Rate (CVR) method. It could be proved in Figure~\ref{fig:cvr} that the CVR method extends the coverage from finite discrete points to the whole continuous range of the R-D curve while R-D performance not degrades.

\subsection{Variable Rate of Hyperprior}

\begin{figure}
	\centering
	\vspace{-0.1cm}
	\includegraphics[width=8.3cm]{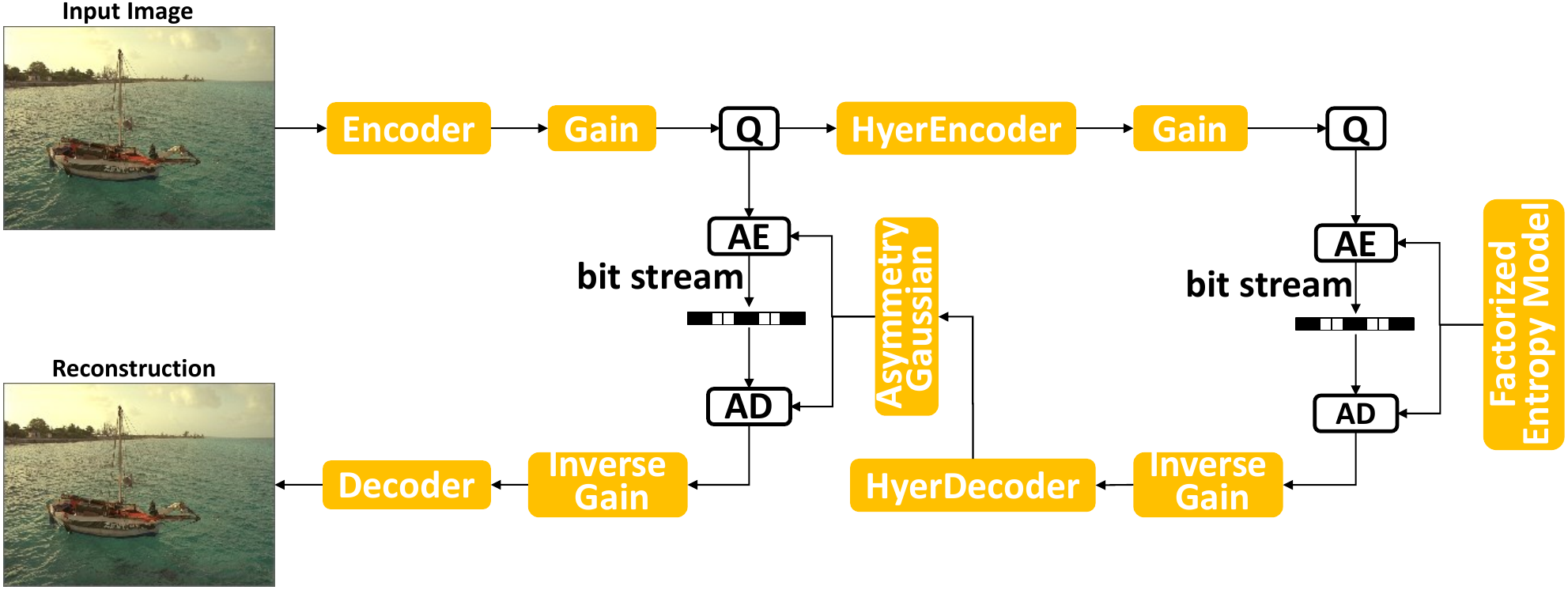}
	\caption{The network architecture of the HCVR. Based on the CVR framework, the HCVR just adds a pair of gain units to the hyperprior autoencoder to obtain flexible entropy estimation.}
	\setlength{\abovecaptionskip}{-0.2cm}
	\vspace{-0.70cm}
	\label{fig:framework_hyper}
\end{figure}

\begin{figure*}[ht]
	\centering
	\vspace{-0.1cm}
	\includegraphics[width=15.3cm]{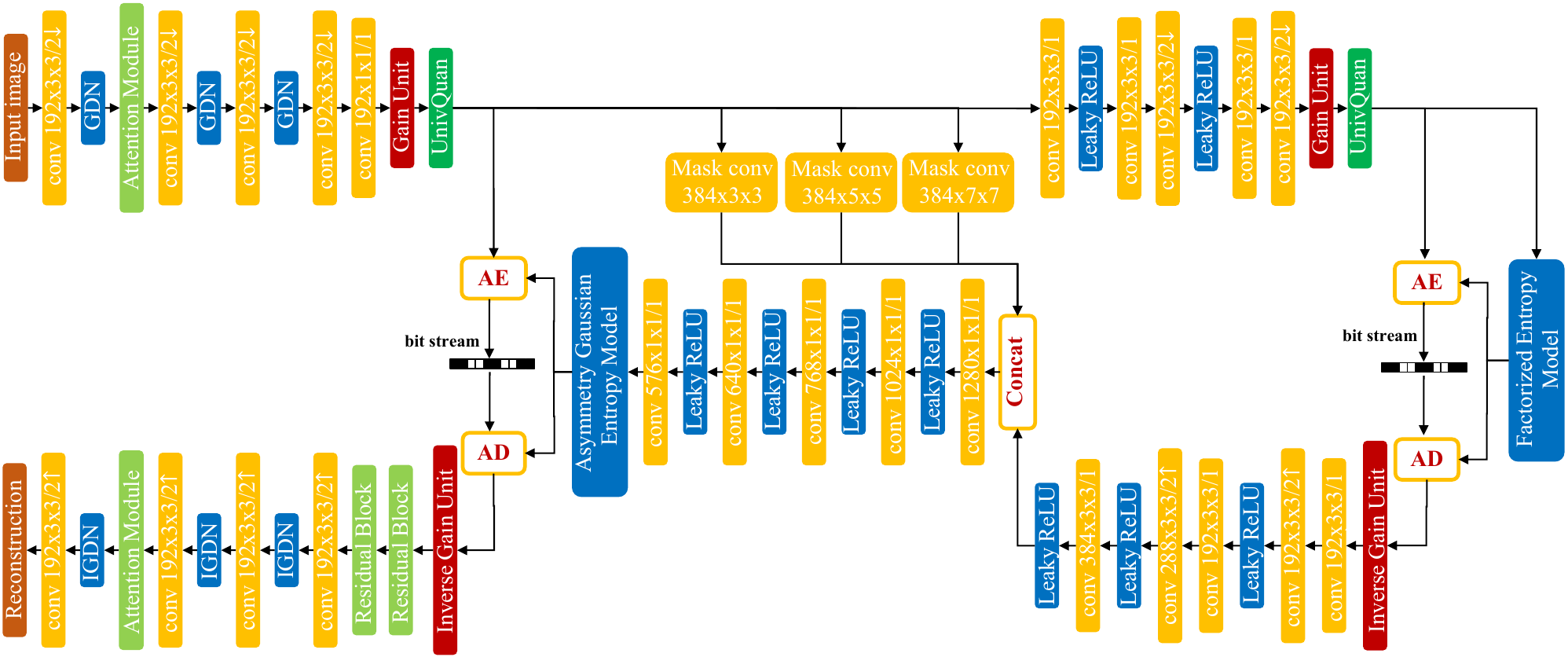}
	\caption{The network architecture of AG-VAE. Convolution parameters are denoted as the number of filters$\times$kernel height$\times$kernel width / stride, where $\uparrow$ and $\downarrow$ represent upsampling and downsampling respectively. GDN and IGDN represent generalized divisive normalization and the inverse counterpart respectively \cite{Balledensity2016}. Attention Module is used to improve network performance \cite{zhang2019}. AE and AD represent the arithmetic encoder and decoder. Masked convolution \cite{Oord2016} is utilized to enhance entropy estimation accuracy. The gain unit and inverse gain unit have been interpreted above to achieve rate adaptation. UnivQuant represents universal quantization \cite{Ziv1985,Zamir1992}.}
	\setlength{\abovecaptionskip}{-0.2cm}
	\vspace{-0.55cm}
	\label{fig:framework_entire}
\end{figure*}
The hyperprior network \cite{BallVariational2018, MinnenJoint2018} can capture the latent representation's spatial dependencies and achieve a more accurate estimation of its distribution. The hyperprior network also adopts the autoencoder structure. It generates the hyperprior latent presentation $z$, which is modelled by a non-parametric, fully factorized entropy model. $z$ also needs to be arithmetically encoded and transmitted, and contributed as a part of the final loss. Therefore, the rate adaption of $z$ helps to reduce the rate of the learned image compression methods containing the hyperprior network. 

The structure of our proposed Hyperprior Continuously Variable Rate (HCVR) method is shown in Figure~\ref{fig:framework_hyper}. Another pair of the gain units are introduced into the hyperprior network to scale the hyperprior $z$. As the hyperprior $z$ can be scaled flexibly, the HCVR method reduces the rate consumption of $z$ without the harm of performance. We will demonstrate the superiority of the HCVR method over the CVR method in the following experiments of Section~\ref{section:sec4.3}.

\subsection{Gaussian Entropy Model}
\begin{figure*}[ht]
	\centering
	\vspace{0.25cm}
	\includegraphics[width=15.3cm]{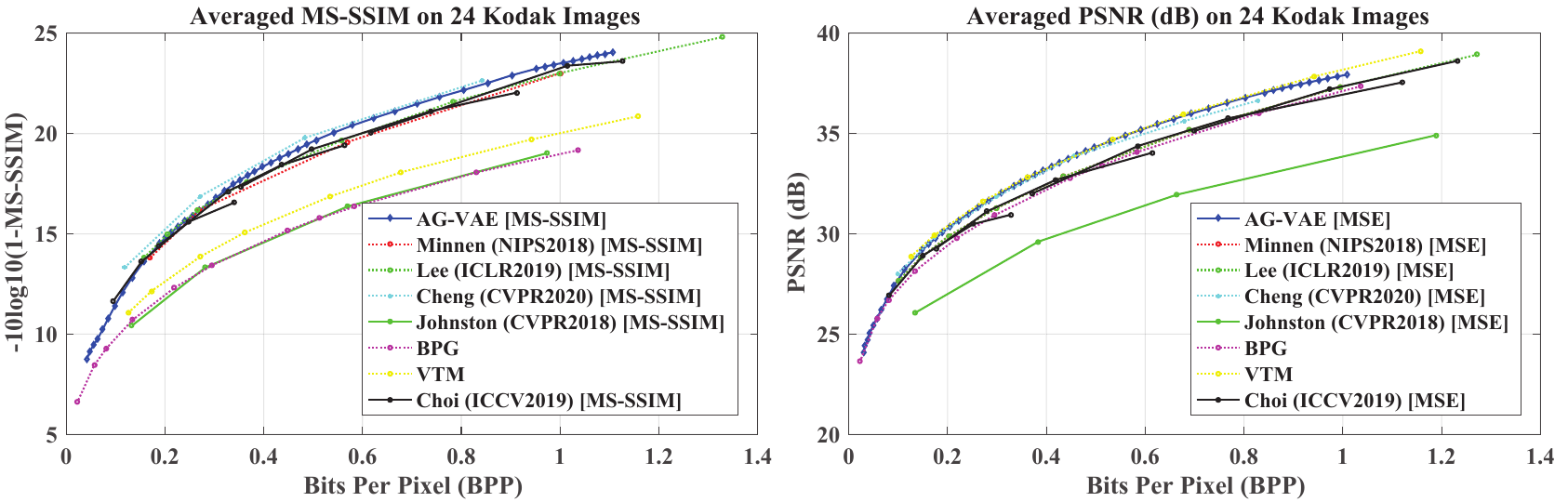}
	\caption{PSNR and MS-SSIM comparison between our variable-rate model AG-VAE and the state-of-the-art image compression methods \cite{Lee2019,MinnenJoint2018,Cheng2020,Choi2019,Johnston2018,Fabrice2014,VTM2019} on 24 Kodak images.}
	\label{fig:best}
	\setlength{\abovecaptionskip}{-0.2cm}
	\vspace{-0.70cm}
\end{figure*}
The matching degree of the parameterized distribution model and real marginal distribution of the latent representation is a significant factor for the expected code length (bit rate) of the quantized latent representation, which decides R-D performance.  As the current mainstream model to estimate the distribution of the latent representation, the mean and scale Gaussian entropy model \cite{BallVariational2018} can be formulated as: 
\begin{equation}
	p_{\hat{y}|\hat{z}}(\hat{y}|\hat{z})\sim N(\mu ,\sigma ^{2}),
	\setlength\abovedisplayskip{3pt}
	\setlength\belowdisplayskip{3pt}
	\label{eq:gaussian}
\end{equation}
where $\mu$ and $\sigma ^{2}$ represent the estimated mean and scale parameters of the latent representation. However, the symmetric Gaussian entropy model has insufficient degrees of freedom and may induce large estimation error for natural images with other distributions. Therefore, we propose the asymmetric Gaussian entropy model \cite{2010asymmetric} as follows:
\begin{equation}
	p_{\hat{y}|\hat{z}}(\hat{y}|\hat{z})\sim N(\mu ,\sigma_{l}^{2},\sigma _{r}^{2}),
	\setlength\abovedisplayskip{3pt}
	\setlength\belowdisplayskip{3pt}
	\label{eq:asymmetric}
\end{equation}
where $\sigma_{l}^{2}$ and $\sigma_{r}^{2}$ represent the estimated left-scale and right-scale parameter of the latent representation. The asymmetric Gaussian model can achieve better entropy estimation for samples, which do not obey the symmetric gaussian distribution strictly. Besides, all the parameter, including $\mu$, $\sigma_{l}^{2}$ and $\sigma_{r}^{2}$, are learnable during the training process so that to the extreme that $\sigma_{l}^{2}$ and $\sigma_{r}^{2}$ are the same, the asymmetric Gaussian model could degrade to the symmetric Gaussian model. Therefore, the proposed asymmetric Gaussian entropy model is more flexible and accurate for entropy estimation of the latent representation. 

\subsection{Network Architecture}
Our image compression framework AG-VAE is depicted in Figure~\ref{fig:framework_entire}. We adopt the network in \cite{MinnenJoint2018} as the basic architecture and introduce gain units to realize continuous rate adaptation. The number of channels of the latent representation $y$ is set to 192 and the kernel size is set to $3 \times 3$. The asymmetric Gaussian entropy model is used to replace the symmetric Gaussian entropy model \cite{MinnenJoint2018} to achieve more accurate entropy estimation. Thus, $192 \times 3$ channels are required for the mean, left-scale, and right-scale parameters of the asymmetric Gaussian entropy estimation. Besides, we also adopt some optimization methods such as the attention module \cite{zhang2019}, Universal quantization \cite{Ziv1985,Zamir1992}, parallel context models \cite{Oord2016}, which have been introduced into the deep image compression methods \cite{ZhouLei2019,Choi2019,ZhouJing2019}, to enhance the R-D performance of the AG-VAE framework.

\section{Experiments}
\subsection{Implemental Details}
\noindent \textbf{Training } The training set consists of a self-building dataset and a training dataset provided in the CLIC2020 \cite{CLIC2020}. The self-building dataset contains $5,000$ high-quality images collected in various scenes. These images are sampled to $2,000 \times 2,000$ pixels and saved as lossless PNGs to avoid compression artifacts. We extract two million patches from these downsampled images with a size of $256 \times 256$ to train the network.
We train the model with Adam optimizer \cite{Diederik2015} for 12 epochs, where the batch size is set to 8, and the learning rate is initially set to $10^{-4}$ and reduced to half at the 6th epoch. In our experiments, $n$ denotes the number of gain vector pairs jointly trained with the AG-VAE framework, which is the same as the number of Lagrange multipliers. We prepare two sets of Lagrange multipliers $B_{msssim}=\left \{0.07, 0.03, 0.007, 0.003, 0.001, 0.0006 \right \}$ and $B_{mse}= \left \{ 0.05, 0.03, 0.007, 0.003, 0.001, 0.0003 \right \}$, which correspond to the models trained with MS-SSIM and MSE loss respectively. In the training process, we randomly select $s$ from 1 to 6 in each iteration to obtain the gain vector $m_{s}$, inverse-gain vector  ${m}'_{s}$ and Lagrange multiplier $\beta_{s}$ from gain matrix $M$, inverse-gain matrix ${M}'$ and $B_{msssim/mse}$. The selected gain/inverse-gain vector will be optimized jointly with the entire framework under the corresponding Lagrange multiplier.

\noindent \textbf{Inference } Given the target image and the target rate, we can obtain large-scale discrete rate adaptation by selecting the index $s$, while adjusting the interpolation coefficient $l$ to achieve fine continuous rate adaptation. The bit rate increases as the values of $s$ and $l$ increase. When $l$ is equal to 0 or 1, the discrete rate at $s$ or $s+1$ can be achieved. In practical use, the parameters $s$ and $l$ are also arithmetically encoded and decoded along with the latent representation.

\subsection{Performace Comprasion}
\noindent \textbf{Rate-distortion Performance } As shown in Figure~\ref{fig:best}, We compare the performance of our variable-rate framework AG-VAE to the state-of-the-art learned image compression methods \cite{Lee2019,MinnenJoint2018,Cheng2020} deploying multiple fixed-rate models, the variable-rate learned image compression methods \cite{Choi2019,Johnston2018}, and the classical image compression codecs \cite{Fabrice2014,VTM2019} on the Kodak dataset \cite{Kodak1992}. The results optimized by MSE or MS-SSIM are presented in two separate plots. With a single model, AG-VAE achieves better R-D performance than those of multiple-networks methods \cite{Lee2019,MinnenJoint2018,Cheng2020} in PSNR, which are believed to be the state-of-the-art ANN-based approach. In MS-SSIM, AG-VAE obtains comparable R-D performance with that of Cheng et al. \cite{Cheng2020} and even better R-D performance than multiple-networks methods \cite{Lee2019,MinnenJoint2018}. Compared with other variable-rate learned image compression methods \cite{Johnston2018,Choi2019}, AG-VAE achieves better R-D performance and adjusts rate flexibly while avoiding performance degradation. In particular, the AG-VAE obtains better results than the widely used classical image codec BPG \cite{Fabrice2014} and yields competitive results with VTM \cite{VTM2019} in PSNR, which is considered to be the best intra-frame encoding methods of the next-generation compression standard Versatile Video Coding (VVC) \cite{ohm2018versatile}.

\begin{figure}
	\centering
	\vspace{-0.1cm}
	\includegraphics[width=7.8cm]{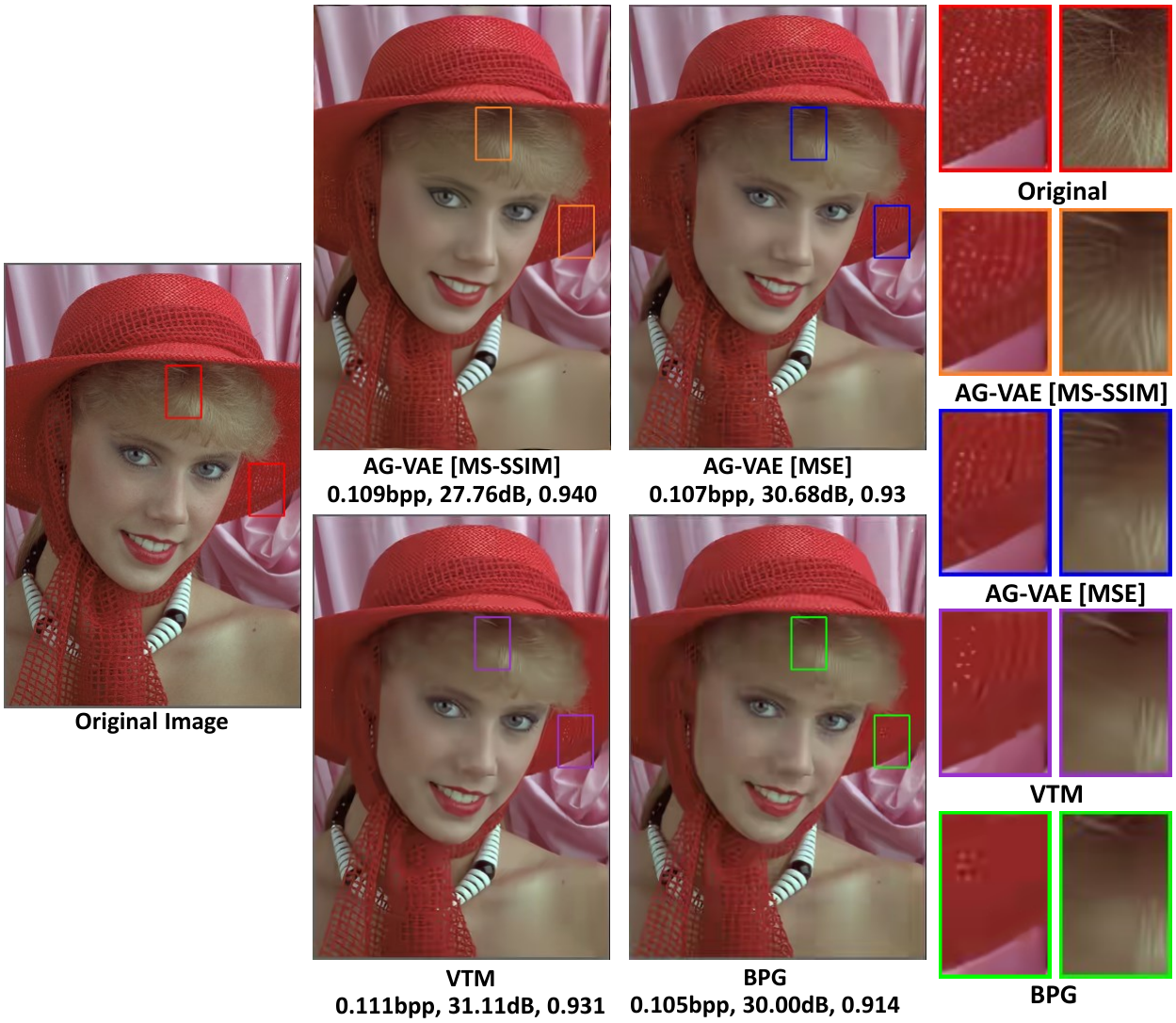}
	\caption{Visualization comparion of reconstructed images $kodim04$ from Kodak dataset with approximately 0.1 bpp.}
	\setlength{\abovecaptionskip}{-0.2cm}
	\vspace{-0.65cm}
	\label{fig:visual}
\end{figure}

\noindent \textbf{Visual Results } Figure~\ref{fig:visual} shows the reconstructed images $kodim04$ with approximately 0.10 bpp from the Kodak dataset \cite{Kodak1992}, which are generated from the AG-VAE methods and classical image compression codecs \cite{Fabrice2014,VTM2019} to assess qualitative performance. We observe that the classical codecs suffer from blurring artifacts \cite{Fabrice2014,VTM2019}. In contrast, the proposed AG-VAE optimized by MSE or MS-SSIM recover more details and alleviate the blurring artifacts better. More qualitative results are included in supplementary materials.

\subsection{Comparison of Variable-Rate Methods} 
\begin{figure}
	\centering
	\vspace{-0.1cm}
	\subfigure{
		\label{fig:vr_method1}
		\includegraphics[width=4.05cm]{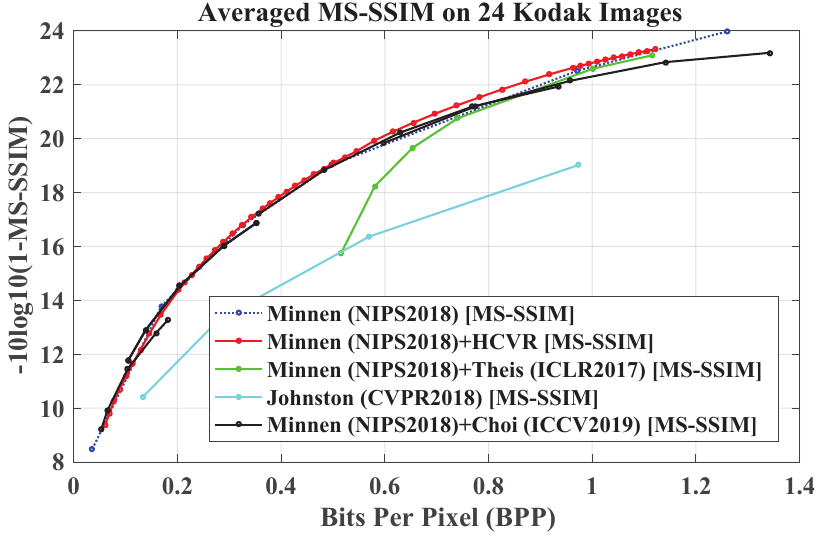}
	}
	\hspace{-3.0mm}
	\subfigure{
		\label{fig:vr_method2}
		\includegraphics[width=4.05cm]{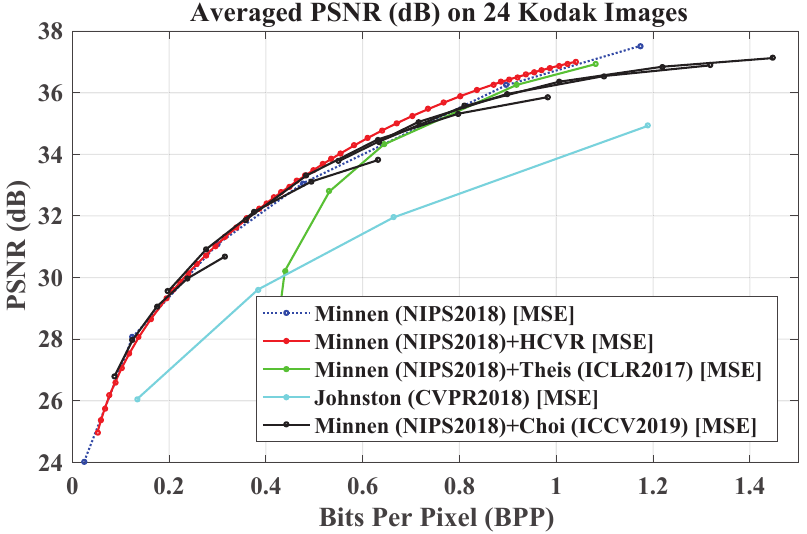}
	}
	\caption{PSNR and MS-SSIM comparison between the proposed HCVR method, \cite{Choi2019}, \cite{TheisLossy2017} and the corresponding method \cite{MinnenJoint2018} with multiple fixed-rate models on 24 Kodak images.}
	\setlength{\abovecaptionskip}{-0.2cm}
	\vspace{-0.45cm}
	\label{fig:vr_method}
\end{figure}

\noindent \textbf{Rate-distortion Performance } To demonstrate the superiority of gain units, we incorporate the proposed HCVR method or previous rate adaptation methods \cite{TheisLossy2017,Choi2019} into the mainstream VAE-based image compression architecture \cite{MinnenJoint2018}. We also compare those methods with the RNN-based variable-rate image compression method \cite{Johnston2018}. From Figure~\ref{fig:vr_method}, we can find that the proposed HCVR method could adjust bit rate flexibly and maintain good performance in the whole range of the R-D curve. However, the method \cite{Choi2019} suffers from performance degradation in high-rate segmentations of the R-D curve and intersection of different bit rate areas of the R-D curve. Meanwhile, the method \cite{TheisLossy2017} suffers from bad performance degradation due to the incompatibility between autoencoder and rate-scaling factors. The RNN-based method \cite{Johnston2018} has far lower R-D performance than counterparts of other methods.

\begin{table}
	\vspace{0.5cm}
	\begin{center}
		\scriptsize
		\begin{tabular}{|m{1.4cm}<{\centering}|m{0.65cm}<{\centering}|m{0.65cm}<{\centering}|m{0.65cm}<{\centering}|m{0.65cm}<{\centering}|m{0.65cm}<{\centering}|m{0.65cm}<{\centering}|}
			\hline
			\multirow{2}*{Basic Method} & \multicolumn{2}{c|}{HCVR} & \multicolumn{2}{c|}{Thesis et al. \cite{TheisLossy2017}} & \multicolumn{2}{c|}{Choi et al. \cite{Choi2019}} \\
			\cline{2-7} 
			& Para. & FLOPs  & Para  & FLOPs & Para.  & FLOPs \\
			\hline 
			Ball$\acute{e}$ et al. \cite{BallVariational2018}  & 0.076 & 0.0004 & 0.023 & 0.0004 & 0.315 & 0.0428 \\ 
			\hline
			Minnen et al. \cite{MinnenJoint2018}  & 0.040 & 0.0002 & 0.010 & 0.0002 & 0.140 & 0.0193 \\ 
			\hline
		\end{tabular}
	\end{center}
    \vspace{-0.3cm}
	\caption{The percentages of additional parameters and computation in the proposed HCVR method, the bottleneck-scaling method \cite{TheisLossy2017}, and the Conditional Conv \cite{Choi2019}.}
	\setlength{\abovecaptionskip}{-0.5cm}
	\vspace{-0.5cm}
	\label{tb1}
\end{table}

\noindent \textbf{Additional Computation and Parameters } Parameter and computational quantity are important metrics of whether the learned image compression methods can be popularized and applied. Variable-rate blocks avoid the multiplication of network memory but introduce new computation modules. The RNN-base method achieves better reconstruction quality with iterations, the running time of which increases proportionally. Therefore, We compare the additional parameter percentages Para. and computation percentages FLOPs between the single fix-rate model \cite{BallVariational2018,MinnenJoint2018} and other classical variable-rate methods, including the proposed HCVR ($n_{hp}=6$), the bottleneck-scaling scheme \cite{TheisLossy2017}, and the Conditional Conv \cite{Choi2019}. It can be seen in Table \ref{tb1} that, compared with the previous classical solution to rate adaptation \cite{Choi2019}, the additional parameter percentages of our HCVR method and the bottleneck-scaling method \cite{TheisLossy2017} are nearly seven times smaller. The additional FLOPs percentages of our HCVR method is nearly 100 times smaller. But the bottleneck-scaling method \cite{TheisLossy2017} suffer from severe performance degradation in the low-rate region. Therefore, we can conclude that the proposed HCVR method utilizes the trivial additional parameters and computation to endow the fixed-rate models with continuous rate adaptation while avoiding performance degradation.

\subsection{Ablation Study}
\begin{figure}
	\centering
	\vspace{0.1cm}
	\subfigure{
		\label{fig:cvr_generic_a}
		\includegraphics[width=8.4cm]{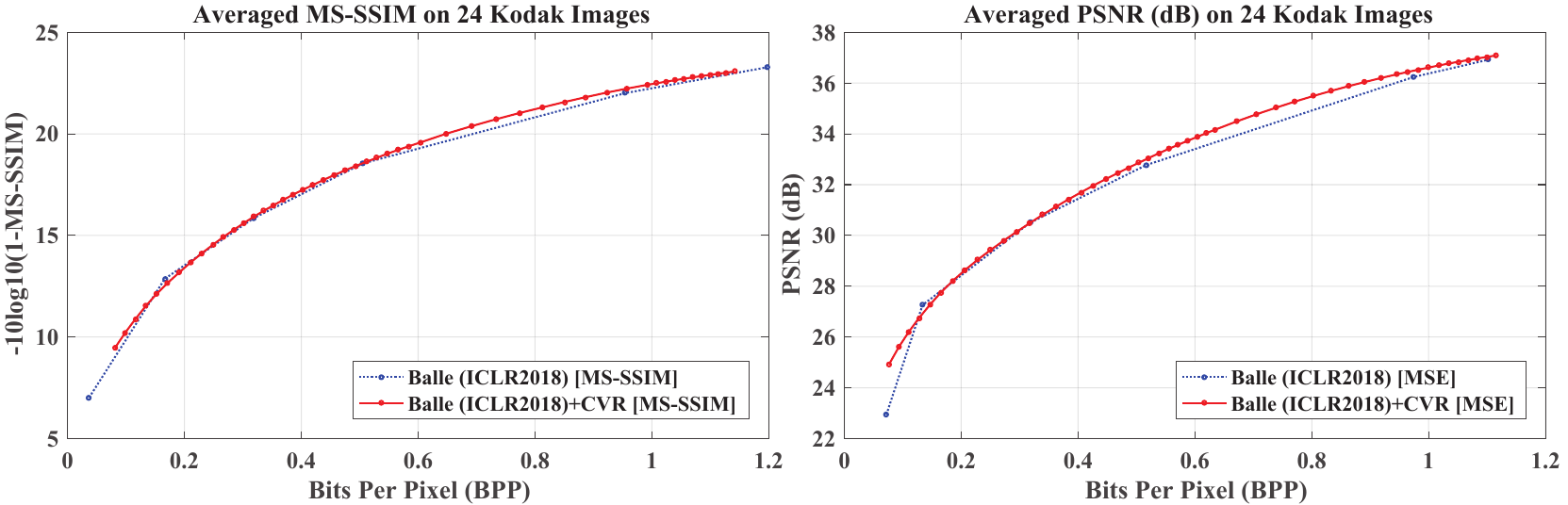}
	}
	\hspace{-2.0cm}
	\subfigure{
		\label{fig:cvr_generic_b}
		\includegraphics[width=8.4cm]{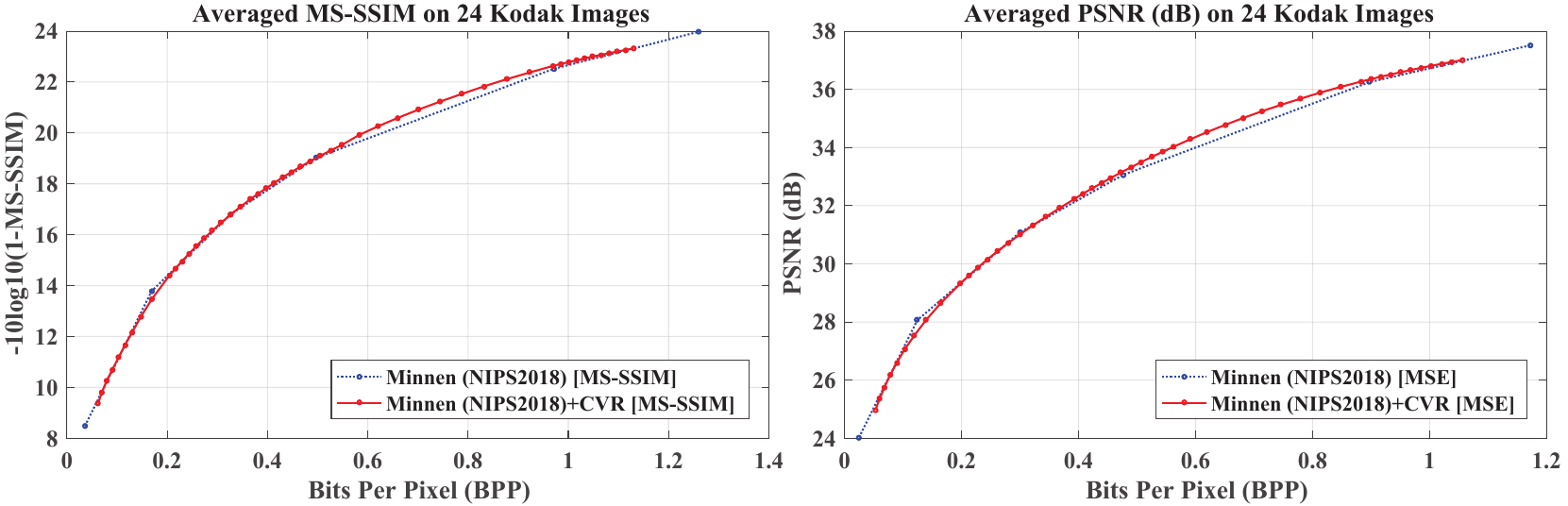}
	}
    \vspace{-0.5cm}
	\caption{PSNR and MS-SSIM comparison between the proposed HCVR and the corresponding learned image compression methods with multiple fixed-rate models on 24 Kodak images. The basic models in the upper row and the bottom row are the method in \cite{BallVariational2018} and the method in \cite{MinnenJoint2018} respectively.}
	\setlength{\abovecaptionskip}{-0.2cm}
	\label{fig:cvr_generic}
	\vspace{-0.50cm}
\end{figure}

\noindent \textbf{Generalizability of Gain Unit } Since there is no need to modify the internal structure of the network, gain units can be easily introduced to almost all the VAE-based image compression methods. We verify the performance of gain units on different VAE-based image compression methods, including Ball$\acute{e}$ et al. \cite{BallVariational2018} and Minnen et al. \cite{MinnenJoint2018}. According to the process introduced in \cite{BallVariational2018, MinnenJoint2018}, we have reproduced the networks, all of which are trained with different Lagrange multipliers separately to get multiple fixed-rate models in different bit rates. Then, we adopt a single model of methods in \cite{BallVariational2018, MinnenJoint2018} as the basic architectures and utilize the CVR method mentioned above to enable them to achieve continuously variable rate in a single model. In Figure~\ref{fig:cvr_generic}, we compare our variable-rate networks with corresponding multiple fixed-rate models in PSNR and MS-SSIM respectively. It could be observed that our variable-rate networks in a single model obtain similar R-D performance with those of the multiple fixed-rate models individually optimized for several discrete fixed Lagrange multipliers. Besides, our variable-rate networks based on the basic architectures \cite{BallVariational2018, MinnenJoint2018} also utilize the exponent interpolation formula to achieve continuous rate adaptation in a single model, like the proposed AG-VAE framework.

\begin{figure}
	\centering
	\vspace{0.4cm}
	\includegraphics[width=8.4cm]{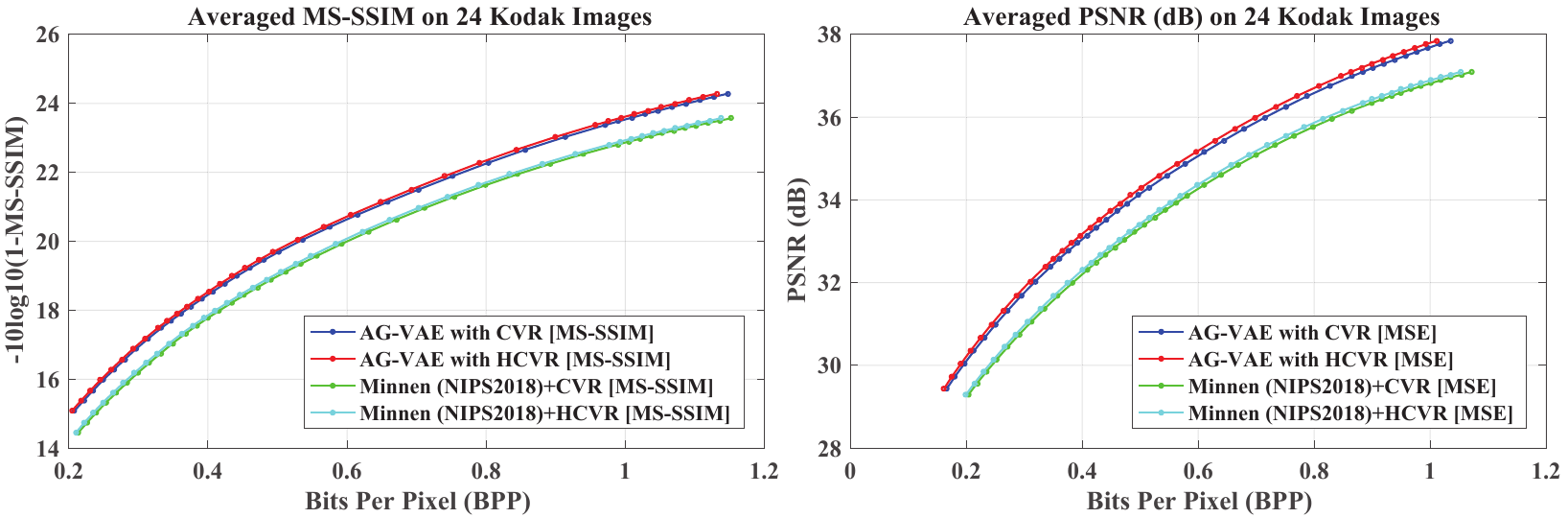}
	\vspace{-0.15cm}
	\caption{PSNR and MS-SSIM comparison between the  HCVR and the  CVR on 24 Kodak images. The basic models are the AG-VAE framework proposed in the paper and the learned image compression method proposed by Minnen et al. \cite{MinnenJoint2018}.}
	\label{fig:hcvr_superiority}
	\setlength{\abovecaptionskip}{-0.2cm}
	\vspace{-0.65cm}
\end{figure}

\noindent \textbf{HCVR Method } By adjusting the bit rate of hyperprior $z$ flexibly, the HCVR method could achieve more accurate entropy estimation for the distribution of the variable-rate latent representation $\hat{y}_{s}$. We adopt the AG-VAE and the image compression method in \cite{MinnenJoint2018} as the basic frameworks to demonstrate the superiority of the HCVR method over the CVR method. As shown in Figure~\ref{fig:hcvr_superiority}, methods with the HCVR method could achieve slightly better R-D performance than the counterpart of methods with the CVR method in the whole bit-rate range. The results demonstrate the effectiveness of the HCVR methods in the learned image compression methods containing the hyperprior network. 

\label{section:sec4.3}

\begin{figure}
	\centering
	\vspace{0.1cm}
	\subfigure{
		\label{fig:rnab_asymmetric1}
		\includegraphics[width=4.05cm]{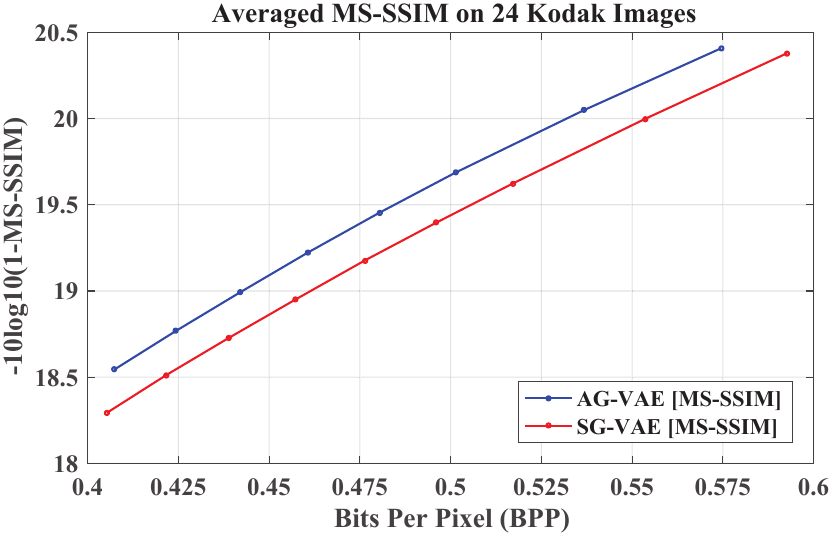}
	}
	\hspace{-3.0mm}
	\subfigure{
		\label{fig:rnab_asymmetric2}
		\includegraphics[width=4.05cm]{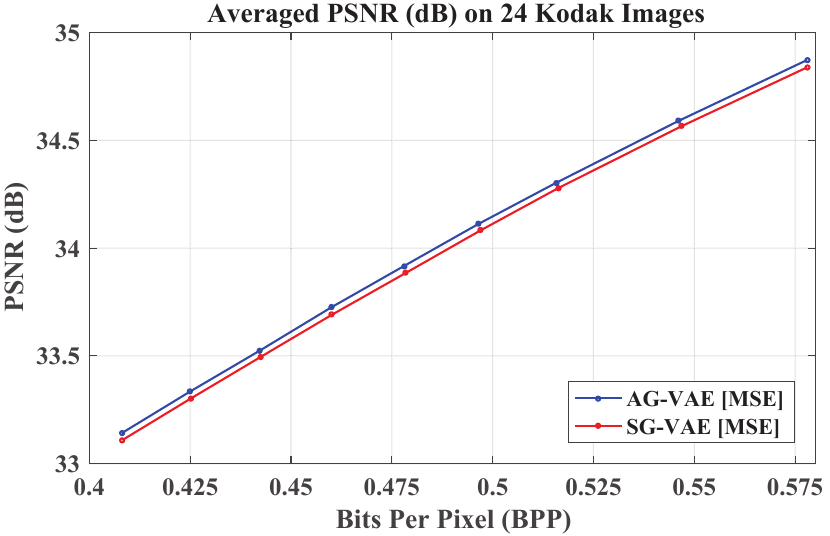}
	}
	\caption{PSNR and MS-SSIM comparison between the AG-VAE and the SG-VAE on 24 Kodak images.}
	\setlength{\abovecaptionskip}{-0.2cm}
	\label{fig:asymmetric_contrast}
	\vspace{-0.50cm}
\end{figure}

\noindent \textbf{Asymmetric Gaussian Model } The images in nature don't always follow a symmetrical Gaussian distribution, which is used to realize entropy estimation in the current learned image compression methods. Therefore, we utilize the asymmetric Gaussian entropy model with a high degree of freedom to reduce entropy estimation errors in the learned image compression methods. When the AG-VAE adopts the previous symmetric Gaussian entropy model, we name it as SG-VAE. To show the difference of compression performance clearly, we cut off R-D curves from $0.4$ to $0.6$ bpp. As shown in Figure~\ref{fig:asymmetric_contrast}, AG-VAE achieves better R-D performance than the counterpart of SG-VAE on both metrics. 

\section{Conclusion}
We propose a novel continuously variable-rate deep image compression framework AG-VAE, which achieves comparable quantitative performance with the SOTA learned image compression methods and even better qualitative performance than the classical image codecs. By utilizing the unevenness of channel redundancy, we design the gain units to achieve discrete rate adaptation while avoiding performance degradation effectively. We then deduce the exponent interpolation to enable gain units to achieve continuous rate adaptation without extra training or modules. From the aspect of additional computation, additional parameters, and performance degradation, gain units are the state-of-the-art solution to rate adaptation for the VAE-based image compression methods. Experimental results demonstrate the effectiveness and efficiency of the gain units with the exponent interpolation. Besides, the proposed asymmetric Gaussian entropy model achieves flexible entropy estimation for raw images, which can also be extended to other learned image compression methods. We also want to utilize the AG-VAE framework on MindSpore \cite{Mindspore}, which is a new deep learning computing framework. These works will be finished in the future.

\bibliographystyle{IEEEbib}
\bibliography{strings}

\end{document}